\documentclass[conference]{IEEEtran}
\IEEEoverridecommandlockouts
\usepackage{cite}
\usepackage{amsmath,amssymb,amsfonts}
\usepackage{graphicx}
\usepackage{tikz}
\usepackage[table]{xcolor}
\usepackage{booktabs}
\usepackage{array}
\usepackage{multirow}
\usepackage{float}
\usepackage{hyperref}
\usepackage{enumitem}
\usetikzlibrary{positioning,arrows.meta,shapes.geometric,calc,fit,backgrounds}
\hypersetup{colorlinks=true,linkcolor=blue,urlcolor=blue,citecolor=blue}

\begin{document}

\title{On the Figures of Merit for Quantum Software Security: Toward a Benchmarking Rubric  \\

\thanks{This work has been supported by the Business Finland through project \textit{SeQuSoS} (Grant No. 112/31/2024), and Research Council of Finland through project \textit{Profi 8 - qSIME} (Grants No. 365343).}
}

\author{\IEEEauthorblockN{Badhon Rahman}
\IEEEauthorblockA{\textit{Faculty of Information Technology} \\
\textit{University of Jyväskylä}\\
Jyväskylä, Finland \\
badhon.b.rahman@jyu.fi}
\and
\IEEEauthorblockN{Majid Haghparast}
\IEEEauthorblockA{\textit{Faculty of Information Technology} \\
\textit{University of Jyväskylä}\\
Jyväskylä, Finland \\
majid.m.haghparast@jyu.fi}
\and
\IEEEauthorblockN{Tommi Mikkonen}
\IEEEauthorblockA{\textit{Faculty of Information Technology} \\
\textit{University of Jyväskylä}\\
Jyväskylä, Finland \\
tommi.j.mikkonen@jyu.fi}

}

\maketitle

\begin{abstract}
Quantum software is increasingly provided through multi-tenant and cloud-based Quantum-as-a-Service (QaaS) stacks. A growing concern about the diverse attack vectors across the pipeline has been demonstrated in recent research. Yet the community has converged on three mature pillars: Scale (Qubit Count), Quality (Quantum Volume), and Speed (Circuit Layer Operations per Second (CLOPS)) for the merit performance figures. Moreover, it has also begun to define software-quality metrics. However, the security of quantum software remains largely unmeasured. A few quantitative security indicators, such as Total Variation Distance (TVD) and Degree of Functional Corruption (DFC), exist. Although they were introduced ad hoc for individual circuit obfuscation techniques, they are incompatible. We assert that the security of quantum software deserves the same attention as the performance: an explicit set of Security Figures of Merit (S-FoMs). The research of this paper is threefold: (i) characterizes a three-layer measurement gap, (ii) proposes a structured S-FoM set organized by ISO/IEC 25010 security sub-characteristics, QaaS pipeline mapping, and measurement maturity, and (iii) defines a benchmarking rubric that normalizes and aggregates S-FoMs into a combined Quantum Software Security Posture (QSSP) score. Additionally, an illustrative reanalysis of published obfuscation techniques has been presented. Our aim is a first step toward security-aware benchmarking of the Quantum Software Stack (QSS). 
\end{abstract}


\section{Introduction}\label{sec:sec1}
Quantum Software is swiftly moving from a theoretical curiosity to engineering practice \cite{b11}. However, its security properties remain far less standardized than its functional or performance characteristics. As quantum programs are progressively executed through the Quantum-as-a-Service (QaaS) platforms, where scheduling, compilation, and multi-tenant hardware remain outside the user's trust boundary and stealthy attacks have already been demonstrated \cite{b13, b20, b10, b19}. The classification of the security challenges of Quantum Software Engineering (QSE) has now been documented \cite{b21}; now we are lacking a way to quantify them. 

A defense is as functional as our ability to measure it. For \textit{performance}, it has standardized terminology, Figures of Merit (FoMs) measured via agreed protocols. For example, Quantum Volume apprehends the largest square circuit that one device runs reliably \cite{b1}. CLOPS frame device capability along speed, quality, and scale as a speed FoM \cite{b2}. Quantum software quality is also emerging, and classical metrics (size and structure measures) are starting to be adapted to quantum programs \cite{b3}. A set of circuit-understandability metrics has been published that aligns with ISO/IEC 25010 \cite{b4, b5}. But for \textit{security}, the quantitative security indicators that are currently available reside in one specific defense. To illustrate, within circuit obfuscation, the Total Variation Distance (TVD) \cite{b6} and the Degree of Functional Corruption (DFC) \cite{b7} measure how well a transformation hides functionality, yet they are not comparable. Therefore, no shared definition or rubric is currently available to permit one practitioner to say that tool \textit{A} is more secure than tool \textit{B} for quantum software.

In this paper, we present \textit{Security Figures of Merit} (S-FoMs) and a rubric that combines them so that security can be measured like performance. Our contributions are as follows: (i) Characterization of three-layer measurement gap among performance FoMs, software-quality metrics, and fragmented/absent security metrics \autoref{sec:sec2}; (ii) For quantum software, a structured set of candidate S-FoMs organized by ISO/IEC 25010 security sub-characteristics, QaaS pipeline mapping, and maturity tag \autoref{sec:sec3}; (iii) a benchmarking rubric that normalizes S-FoMs and aggregates into a composite Quantum Software Security Posture (QSSP) score \autoref{sec:sec4}; and (iv) an illustrative reanalysis of published obfuscation techniques substantiating non-comparability of current reporting and the comparability our rubric restores \autoref{sec:sec5}. In \autoref{sec:sec6}, we present a conclusion and discuss our future work.

\section{Background \& Measurement Gap}\label{sec:sec2}
\subsection{Layer 1: Performance Figures of Merit (Mature)}
For quantum computers, benchmarks are defined by FoMs or key performance indicators and measured under a stated protocol. Component level FoMs encompass coherence times ($T_1$/$T_2$), single-qubit, two-qubit gates, and readout fidelities. In addition, the system level FoMs aggregate Quantum Volume \cite{b1} and the speed / quality / scale with CLOPS \cite{b2}. The application level FoMs mainly report success on the real workloads. This layer is proportionately standardized and supported by tools. As a result, provides us with a template with a metric, a protocol, and a normalization that makes devices comparable.

\subsection{Layer 2: Software Quality Metrics (Emerging)}
Quantum Software Engineering (QSE) has started to import software measurements into quantum programs. Various research, for example, Zhao \cite{b3} adapted structure and size metrics, including lines of code, Halstead, cyclomatic complexity, and information flow, to the quantum software. In another research work, proposed and empirically validated metrics for quantum circuit complexity and understandability aligning with the ISO/IEC 25010 quality model and supported by an automatic computation tool (QMetrics) \cite{b4}, \cite{b5}. Primarily, these efforts are operationalized in \textit{understandability} and \textit{maintainability} but not \textit{security}. However, security holds a place in the nine characteristics of ISO/IEC 25010:2023, and sub-characteristics contain integrity, confidentiality, non-repudiation, accountability, authenticity, and resistance.

\subsection{Layer 3: Security Metrics (Fragmented / Absent)}
Circuit obfuscation is the only domain in the security level that is frequently assessed against the untrusted compilers. TVD quantifies the statistical distance of the original and obfuscated output distributions \cite{b6}. The higher the value of TVD, indicates stronger the obfuscation.  In contrast, DFC measures the functional corruption on [-1, 1], and here lower value of DFC means greater concealment \cite{b7}. A methodology called "Split Compilation" was presented to keep the intellectual properties (IPs) secure from untrusted compilers by splitting the quantum circuit into different parts and sending them to multiple compilers at different times \cite{b8}. Additionally, a secure compilation methodology of quantum circuits for untrustworthy compilers by embedding one small random circuit was shown in recent research \cite{b9}. However, all these available indicators are valuable but fragmented. Each one of them is linked to one technique; none is crafted as a reusable FoM. Likewise, the other wider threat surface, like authentication, multi-tenant isolation, and accountability, has no comparable metric. At present, \textit{post-quantum cryptography} benchmarking surfaces as "quantum security metrics" and evaluates mainly the cryptographic strength of quantum algorithms. It is a completely different layer from the security of the QSS. Hence, the gap is specific, and there is no shared, multi-property set of S-FoMs and rubric to compare quantum software tools on security.

\section{Security Figures of Merit For Quantum Software}\label{sec:sec3}
We define the \textit{Security Figures of Merit} for quantum software as a quantity that (a) measures one security characteristic or overhead experienced by a quantum software artifact or its execution, (b) are calculated by a stated protocol against one declared attack model, and (c) can be normalized to a common [0,1] scale; here, a higher score denotes stronger security. In Table \ref{tab:fomc}, we propose a set of candidates organized by ISO/IEC 25010:2023 security sub-characteristics with a cross-cutting overhead row. In addition, the table also mapped with the the QaaS pipeline stages of \cite{b15} (\textbf{S1}: Developer Environment, \textbf{S2}: Authentication \& Submission, \textbf{S3}: Cloud Orchestration \& Compilation, \textbf{S4} - Quantum Hardware Execution, \textbf{S5}: Result Return Path, \textbf{S6}: Hybrid Iteration Loop), and labeled with measurement maturity: E = Established (Existing and defined metric or protocol), A = Adaptable (Existing measurable quantity), and P = Proposed (no quantum-specific quantity yet).

\begin{table*}[!t]
\centering
\caption{Characteristics Security Figures of Merit for quantum software security corresponding by ISO/IEC 25010:2023 security sub-characteristics, aligning with QaaS pipeline stages, and elucidated with measurement maturity (E: Established, A: Adaptable, P: Proposed).}
\label{tab:fomc}
\renewcommand{\arraystretch}{1.20}
\footnotesize
\begin{tabular}{@{}l l p{6.4cm} c c l@{}}
\toprule
\textbf{Property} & \textbf{Security Figure of Merit} & \textbf{What it quantifies} & \textbf{Stage} & \textbf{Mat.} & \textbf{Basis} \\
\midrule
\multirow{3}{*}{Confidentiality}
 & IP obfuscation strength (OS) & Statistical distance (Obfuscation Quality) between original and obfuscated output (TVD) & S3 & E & \cite{b6,b7} \\
 & Functional concealment (FC) & Degree of Functional Corruption (DFC) $\in[-1,1]$; lower is stronger & S3 & E & \cite{b7} \\
 & Reverse-engineering resistance (RER) & Adversary effort/success to recover circuit or compiler IP from circuit snapshots or passes & S3 & A & \cite{b8,b12} \\
 & Model-extraction resistance (MER) & Clone accuracy (comparison with the original) by an QML-model reverse-engineering extraction adversary; lower is stronger & S5,S6 & A & \cite{b16} \\
\midrule
\multirow{2}{*}{Integrity}
 & Result tamper detectability (RTD) & Detectability of adversarial result/parameter tampering via cross-device shot and iteration distribution & S5,S6 & A & \cite{b17} \\
 & Output verifiability (OV) & Accreditation based upper bound on returned output deviation from ideal & S4,S5 & A & \cite{b18} \\
\midrule
Authenticity
 & Device attestation strength (DAS) & Quantum existential unforgeability (qPUF with classical-PUF fingerprint uniqueness) & S2,S4 & A & \cite{b22} \\
\midrule
\multirow{2}{*}{\parbox{2.1cm}{Non-repudiation \& accountability}}
 & Provenance completeness (PC) & Fraction of pipeline events carrying attributable, tamper-evident records & S2--S6 & P & \cite{b15} \\
 & Audit-log tamper-evidence (ALT) & Cryptographic integrity coverage of job logs (e.g., hash-chained) & S2--S6 & P & --- \\
\midrule
\multirow{3}{*}{Resistance}
 & Crosstalk susceptibility (XS) & Inter-circuit interference under co-tenancy, lower is stronger & S4 & A & \cite{b23,b14} \\
 & Co-tenancy output divergence (COD) & Hellinger distance between output distributions of co-located executions & S4 & A & \cite{b24} \\
 & Attack success rate (ASR) & Fraction of injection/crosstalk attacks succeeding under a defense; lower is stronger & S4 & A & \cite{b10,b19, b24} \\
\midrule
Cross-cutting
 & Security overhead (SO) & Added depth/gate/runtime/shot cost incurred by the controls & S1--S6 & E & \cite{b7,b8,b17} \\
\bottomrule
\end{tabular}
\end{table*}

We have observed two essential points from the table. Firstly, confidentiality is the sole well-instrumented property (obfuscation FoMs) and confirms that at present quantitative security research is concentrated at S3. Secondly, the accountability row (PC \& ALT) labeled with P is the least explored property without an existing quantum-specific metric. In addition to the maturity column, the empty cell resembles the place where new protocols are urgently needed.

\section{Benchmarking Rubric}\label{sec:sec4}
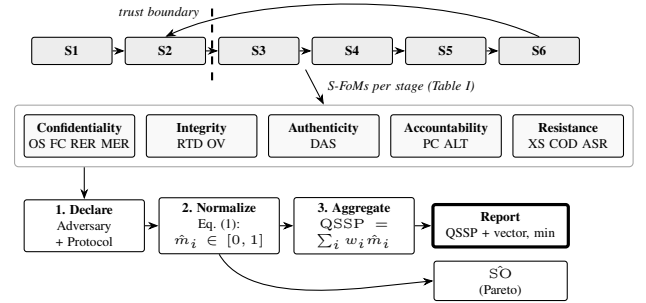
\begin{figure}[t]
\centering

\begin{tikzpicture}[font=\tiny,
  stage/.style={draw, fill=gray!14, rounded corners=1pt, inner sep=1.6pt, align=center, minimum height=3.6mm, text width=9.5mm},
  chip/.style={draw, fill=gray!5, rounded corners=1pt, inner sep=1.6pt, align=center, text width=13.5mm, minimum height=5.6mm},
  rstep/.style={draw, rounded corners=1pt, inner sep=2pt, align=center, text width=14.5mm, minimum height=5.6mm},
  outb/.style={draw, very thick, rounded corners=1pt, inner sep=2pt, align=center, text width=16.5mm, minimum height=5.6mm},
  arr/.style={-{Stealth[length=1.4mm]}, thin}]

\node[stage] (s1) {\textbf{S1} };
\node[stage, right=1.6mm of s1] (s2) {\textbf{S2} };
\node[stage, right=1.6mm of s2] (s3) {\textbf{S3} };
\node[stage, right=1.6mm of s3] (s4) {\textbf{S4} };
\node[stage, right=1.6mm of s4] (s5) {\textbf{S5} };
\node[stage, right=1.6mm of s5] (s6) {\textbf{S6} };
\foreach \a/\b in {s1/s2,s2/s3,s3/s4,s4/s5,s5/s6}{\draw[arr] (\a)--(\b);}
\draw[arr] (s6.north) to[out=155,in=25,looseness=0.7] (s2.north);

\draw[dashed, thick] ($(s2.east)!0.5!(s3.west)+(0,5.6mm)$) -- ($(s2.east)!0.5!(s3.west)-(0,4.6mm)$);
\node[anchor=east, font=\tiny\itshape] at ($(s2.east)!0.5!(s3.west)+(-0.5mm,5.0mm)$) {trust boundary};

\node[chip, below=6.5mm of s1.south west, anchor=north west, xshift=-1mm] (c1) {\textbf{Confidentiality}\\ OS FC RER MER};
\node[chip, right=1.4mm of c1] (c2) {\textbf{Integrity}\\ RTD OV};
\node[chip, right=1.4mm of c2] (c3) {\textbf{Authenticity}\\ DAS};
\node[chip, right=1.4mm of c3] (c4) {\textbf{Accountability}\\ PC ALT};
\node[chip, right=1.4mm of c4] (c5) {\textbf{Resistance}\\ XS COD ASR};
\node[draw=gray!70, thin, rounded corners=1.5pt, inner sep=1.2mm, fit=(c1)(c5)] (band) {};
\draw[arr] ($(s3.south)!0.5!(s4.south)+(0,-0.6mm)$) -- node[right=0.5mm, font=\tiny\itshape]{S-FoMs per stage (Table~I)} (band.north);

\node[rstep, below=5.5mm of c1.south west, anchor=north west] (r1) {\textbf{1. Declare}\\ Adversary + Protocol};
\node[rstep, right=1.8mm of r1] (r2) {\textbf{2. Normalize}\\ Eq.~(1): $\hat m_i\in[0,1]$};
\node[rstep, right=1.8mm of r2] (r3) {\textbf{3. Aggregate}\\ $\mathrm{QSSP}=\sum_i w_i \hat m_i$};
\node[outb, right=2.2mm of r3] (r4) {\textbf{Report}\\ QSSP + vector, min};
\node[draw, rounded corners=1pt, inner sep=1.6pt, align=center, text width=16.5mm, below=1.6mm of r4] (so) {$\hat{\mathrm{SO}}$ \\ (Pareto)};
\draw[arr] (band.south -| r1) -- (r1.north);
\draw[arr] (r1)--(r2); \draw[arr] (r2)--(r3); \draw[arr] (r3)--(r4);
\draw[arr] (r2.south) to[out=-60,in=185,looseness=0.75] (so.west);
\end{tikzpicture}

\caption{Three step overview of our research: Six stage QaaS pipeline outlined with the S-FoM candidates organized with standard security sub-characteristics and combines them with the rubric into QSSP.}

\label{fig:ro}
\end{figure}

A set of S-FoMs develops into a benchmark only when the metrics are commensurable and integrated under explicit assumptions. Here, we propose three steps that are designed with the classical benchmark requirements: relevance, reproducibility, fairness, verifiability, and usability \cite{b26}.\\
\textbf{Attack-model declaration:} Each and every S-FoM will be reported against one stated adversary. A security number without an adversary is meaningless. For instance, obfuscation strength is conditional on the reverse-engineering attack keeping in mind. One defense that is strong against one specific attack may fail against another. Hence, the rubric requires a one-line threat model (capabilities, access, and stage) per reported S-FoM. So that, all together with the measurement protocol (backend, calibration snapshot, workload suite, shot budget, seed) makes the number reproducible. \\
\textbf{Normalization:} For each S-FoM value $m_i$ is mapped to an orientation-corrected score $\hat{m}_i\in[0,1]$, here 1 is strongest security. The score is calculated by max-min scaling between two declared reference points. The weakest value $m_i^{\circ}$ and a strongest or saturation value $m_i^{\star}$ \cite{b27}: 
\begin{equation}
    \hat{m}_i=\min\bigl\{1,\;\max\bigl\{0,\;(m_i-m_i^{\circ})/(m_i^{\star}-m_i^{\circ})\bigr\}\bigr\}
\end{equation}
The reference absorb orientation ($m_i^{\star} < m_i^{\circ}$ for lower is stronger metrics such as DFC, MER, XS, ASR, COD, and SO). For example, DFC applies ($m_i^{\star} = -1$ , $m_i^{\circ} = 1$) and force unbounded quantities (grows to infinity) such as adversary effort to declare a saturation point. They are part of the benchmark definition and are published with the scores. \\
\textbf{Weighted aggregation:} A composite Quantum Software Security Posture score is a weighted sum over the chosen S-FoMs: 
\begin{equation}
    \mathrm{QSSP} = \sum_{i=1}^{N} w_i\,\hat{m}_i,\qquad \sum_{i} w_i = 1,\; w_i\ge 0,
\end{equation}
Additionally, these weights are elicited by a multi-criteria method such as Analytic Hierarchy Process (AHP) \cite{b25}. In order that conflicting and context-dependent priorities (e.g., IP secrecy) are made clear rather than hidden. The methodology of composite indicator includes two requirements \cite{b27}: i) One weighted sum is completely compensatory; one property's strength can mask collapse on another. At times, it can happen with the weakest-link character of security. For this reason, QSSP is always published with a score vector and its minimum with the geometric form ($\prod_i \hat{m}_i^{\,w_i}$) as the conservative alternative. ii) The reported rankings must be demonstrated to plausible weight perturbations. Since security is purchased with resources, the Security Overhead ($\hat{\mathrm{SO}}$) score is not incorporated into QSSP, but is stated alongside it. A tool that dominates another on both the QSSP and the overhead axes is preferred, analogous to the Pareto-frontier reading of the volumetric performance benchmark \cite{b28}. Here, a number is needed and $\mathrm{QSSP}_\lambda=\mathrm{QSSP}\,(1-\lambda(1-\hat{\mathrm{SO}}))$ reveals the trade-off through a tunable $\lambda$. The \autoref{fig:ro} summarizes our research by dividing it into three steps: 1. QaaS Pipeline, 2. Security sub-characteristics, and 3. The Rubric.

\section{Illustrative Reanalysis}\label{sec:sec5}
To demonstrate both the concern and the solution, we reanalyze two published quantum software security controls (confidentiality) at the Cloud Orchestration and Compilation level. The values are taken from the original papers and not new measurements. The Dummy CNOT gate insertion technique \cite{b6} reports TVD up to 28.83\% with modest gate and depth overhead. In contrast, the randomized reversible gate technique \cite{b7} reports 1.92 and uniformly negative DFC with 1-3\% degradation in fidelity. However, obfuscation literature including these two reports "TVD" under incompatible definitions. Such as normalizing the summed $\ell_1$ count difference by the number of shots varying in [0,2], while the textbook TVD divides by twice the shots and varies in [0,1]. Two different "TVD" values defined on different scales cannot be compared. This is exactly the motivation for standardized S-FoM definitions.  

When a single TVD definition and common attack model are fixed, then the OS, FC, and RER scores are normalized by equation (1). Scores will be combined with weights reflecting an IP protection priority on a multi-tenant cloud. As a result, the rubric will get a comparable confidentiality sub-score for each of the tool. By reporting $\hat{\mathrm{SO}}$ individually, makes the security overhead tradeoff explicit. Ultimately, the contribution is not the ranking itself but that structured S-FoMs turn non-comparable and scattered numbers into an interpretable comparison.

\section{Conclusion and Future Work}\label{sec:sec6}
We have argued that quantum software security should be measured with the same precision as performance and proposed Security Figures of Merit organized and structured by ISO/IEC 25010, QaaS pipeline, and maturity tag. In addition, a rubric is defined that normalizes and aggregates them into a composite posture score with a security overhead tradeoff. A reanalysis of the published obfuscation techniques indicated that current reporting is non-comparable, and S-FoMs restore the compatibility. In our next step, we plan to design and develop a benchmarking tool that computes the established S-FoMs over Qiskit and Cirq programs, together with the protocols for the metrics that are currently unmeasured.

\end{document}